# Quantum mechanical complementarity probed in a closed-loop Aharonov-Bohm interferometer

Dong-In Chang[1], Gyong Luck Khym[1,2], Kicheon Kang[2], Yunchul Chung[3], Hu-Jong Lee[1,4], Minky Seo[3], Moty Heiblum[5], Diana Mahalu[5], Vladimir Umansky[5]

[1]*Department of Physics, Pohang University of Science and Technology, Pohang 790-784, Republic of Korea*

[2]*Department of Physics, Chonnam National University, Gwangju 500-757, Republic of Korea*

[3]*Department of Physics, Pusan National University, Busan 609-735, Republic of Korea*

[4]*National Center for Nanomaterials Technology, Pohang 790-784, Republic of Korea*

[5]*Braun Center for Submicron Research, Department of Condensed Matter Physics, Weizmann Institute of Science, Rehovot 76100, Israel*

**According to Bohr's complementarity principle[1], a particle possesses wave-like properties only when the different paths the particle may take are indistinguishable. In a canonical example of a two-path interferometer with a which-path detector, observation of interference and obtaining which-path information are mutually exclusive[2,3]. Such duality has been demonstrated in optics with a pair of correlated photons[4] and in solid-state devices with phase-coherent electrons[5]. In the latter case, which-path information was provided by a charge detector embedded near one path of a two-path electron interferometer[5]. Note that suppression of interference can always be understood either as obtaining path information or as unavoidable back action by the detector[3]. The present study reports on dephasing of an Aharonov-Bohm (AB) ring interferometer[6] via a**

**coupled charge detector adjacent to the ring. In contrast to the two-path interferometer, charge detection in the ring does not always provide path information. Indeed, we found that the interference was suppressed only when path information could be acquired, even if only in principle. This demonstrates that dephasing does not always take place by coupling the 'environment' to the interfering particle: path information of the particle must be available too. Moreover, this is valid regardless of the strength of environment-interferometer coupling, which refutes the general notion of the effect of strong interaction with the environment[7]. In other words, it verifies that an acquisition of which-path information is more fundamental than the back-action in understanding quantum mechanical complementarity.**

Recently, a series of electronic 'which-path' experiments have been performed in mesoscopic solid-state devices.[5] The devices, fabricated in the plane of a high-mobility two-dimensional electron gas (2DEG), were based on a double-path interferometer, consisting of an *open* Aharonov-Bohm (AB) ring, with a source and a drain of electrons weakly coupled to the open ring[5]. In one path of the interferometer a coherent quantum dot (QD) was embedded[5-6,8], being electrostatically coupled to a quantum-point-contact (QPC) charge detector (in the immediate proximity to the QD). An electron trapped in the QD modified the conductance of the nearby QPC and thus allowed charge detection by the QPC[5,9-11]. Being an open geometry, with multiple grounded drains (bases) along the paths of the electron, assured that only two paths interfered while the backscattered electrons were drained out by the grounded bases. Thus, the detection of a charge inside the QD (by the QPC) provided path information, which led to the suppression of the AB interference oscillations.

In our present study, we employed a '*closed-loop*' AB interferometer[6,12], as shown in Fig. 1a, with a QD and a QPC detector placed in a similar manner as in Buks *et al*[3]. However, in contrast to the previous schemes, the closed geometry allowed an electron to encircle the interferometer loop many times before it reached the drain, making the interferometer an analog of the Fabry-Perot interferometer, where in the closed ring the forward propagating and the backward propagating paths are spatially separated.   Let's look at a couple of examples: Among various possible electron trajectories contributing to the interference, the most probable trajectories, which lead to source-drain conductance oscillations with periods of one (*h/e*; first harmonic) and a half (*h/2e*; second harmonic) flux quantum, are illustrated in Figs. 1b and 1c, respectively[13-14].   In principle, there are infinite number of other possible trajectories that give rise to the first- and the second-harmonic interferences.   However, as will be discussed below, these two sets of trajectories (shown in Figs. 1b and 1c) are the dominant ones.   For the trajectories plotted in Fig. 1b, one can acquire path information by detecting the presence of an added electron inside the QD.   Alternatively, for the trajectories plotted in Fig. 1c, charge detection in the QD does not, in general, provide path information to distinguish between the *blue* and *red* trajectories; since both pass the QD once[12]. Hence, one would expect strong dephasing of the first AB harmonic and not of the second harmonic.   Adding an element of time in the detection process may distinguish between trajectories of an electron trapped in the QD, and suppress the second harmonics also - as we describe below - making this experiment clearly contrasted to its predecessors.

Our closed-loop interferometer (Fig. 1a) was fabricated in a GaAs/AlGaAs heterojunction wafer containing a 2DEG residing about 80 nm below the surface. The electron density was $n_s=1.8\times10^{11}$ cm$^{-2}$ with mobility of $\mu=3.3\times10^6$ cm$^2$V$^{-1}$s$^{-1}$ at 4.2 K, resulting in an elastic mean free path of $l_e \sim 20$ µm. Figure 1a shows a scanning electron micrograph of the device used in this study. Negative voltages were applied to the gates so as to pinch off the 2DEG underneath them. Four side gates ($M_1$, $M_2$, $M_3$, $M_4$) together with an island gate (P) defined an AB ring, with approximately five conducting channels ($N>5$) in each path. The geometrical radius of the interferometer was about 550 nm, which was also confirmed by the ensuing AB interference oscillations. Two QPC's ($Q_1$, $Q_2$), at the source and the drain were used to configure the two-terminal measurement. A QD and a QPC detector were placed to detect charges trapped inside the QD. Two gates ($F_1$, $F_2$) defined the QD and separated it from the QPC detector. The measurements were made by applying a 10 µV RMS excitation voltage to the source and monitoring the output current at the drain (the electron temperature was 140 mK).

In order to examine the characteristics of the QD, the left path was pinched off by applying a large negative voltage to the side gates $M_1$ and $M_2$. The Coulomb blockaded (CB) conductance peaks of the QD were monitored by varying the voltage on the center island gate (supplied by the *air bridge* P, see Figs. 1a and 2a). The QD was tuned to the CB conductance peak marked by the vertical arrow in Fig. 2a. The transmission probability of the QPC detector, defined as $T_d=G_d/(2e^2/h)$, was set to $T_d = 0.1775$, where the dephasing rate was found to be the highest[5]. The close proximity between the QPC detector and the QD lead to a strong modification of its transmission, $\Delta T_d$, as illustrated in Fig. 2a, each time the number of electrons inside QD changed by one. Equivalently, the mere possibility of measuring the added charge to the dot, leads, by the inadvertent back action of the detector on the QD, to dephasing of the dwelling electron in the QD. Following the above-described tuning of the dot, the current through the detector was shut off and the left arm of the interferometer was opened. The interferometer was then tuned to exhibit both the first and the second harmonic of the AB interference.

The effect of the charge (or path) detection on the interference pattern, at different detector bias voltages (0 to 400 µV), is shown in Fig. 2b. The AB oscillations exhibit clearly two harmonics (with the field periodicity of ~4.6 and ~2.3 mT), which are being suppressed with increasing the bias of the detector; as expected. As the number of 'detecting electrons' that pass the detector during the dwelling of the electrons in the QD increases, dephasing is enhanced. Looking at the inset of Fig. 2b, it is clear that the first AB harmonic is much more sensitive to the bias of the detector than the second harmonic. As alluded above the difference arises from the fact that the charge detection cannot distinguish between the two types of trajectories illustrated in Fig. 1c leading to the second harmonic.

To be more specific, there are two major sets of trajectories for the second harmonic. In the first set of trajectories shown in Fig. 1c, the partial wave of an electron starting with the left path at the source makes one and a half clockwise turns around the interferometer (blue); while that starting with the right path makes just a half turn (red). In the second set of trajectories (not shown), the direction of the two partial waves are switched: The shorter path goes though the left and the counterclockwise path makes one and a half turn passing through the QD twice. In the latter case, path information is obtained via charge detection because only the longer path passes through the QD (even twice) while the other path never passes through the

QD. The transmission probability through the QD, $T_P$, can be estimated as follows: since the conductance peak is about $G_{QD}/(2e^2/h) \sim 0.25$ and the number of transverse channels in each path is about $N \sim 5$, the transmission probability is $T_P = G_{QD}/(2e^2/h)/N \sim 0.05$. Therefore, other more complicated sets of trajectories can be ignored due to the low transmission probability through the QD. This also explains the absence of higher-order harmonics with $n>2$. Furthermore, the second set of trajectories, compared to the first set shown in Fig. 1c, can also be neglected due to the large dwell time in the QD ($t_d \sim 2$ ns), which is much larger than the characteristic mean time interval, $t_f$, between successive injection of electrons through the left (shorter) path. From the bias voltage of the interferometer, $V=10$ μV, one finds that $t_f \sim h/2eV \sim 0.2$ ns. For the first set of trajectories shown in Fig. 1c, both the partial waves pass the QD only once, and there is no considerable time delay in arrivals of the two wavepackets. However, for the second set of trajectories, the characteristic time of the shorter path is about $t_f \sim 0.2$ ns and that of the longer path is about $2t_d \sim 4$ ns. This large time delay strongly suppresses the interference due to the lack of overlap of the wave packets taking two different paths. Therefore, the second harmonic is dominated by the trajectories shown in Fig. 1c, which do not provide path information in charge detection process.

One should note that the sets of electron trajectories like those described in Fig. 1c may be distinguished via charge detection in the QD if the difference in their dwell time in the interferometer can be differentiated; hence dephasing also the second AB harmonic. The difference in the paths' length of the two possible trajectories shown in Fig. 1c, going from source to the QD, is about 1.7 μm, and for a Fermi velocity of

$v_F$~$1.62\times10^5$ m/s one gets a time difference ~10 ps. This must be compared to the time difference between consecutive electron arrivals in the QPC detector, $h/2eV_d$, where $V_d$ is the applied bias voltage to the QPC detector[5]. This simple argument leads to a conclusion that path information for the trajectories shown in Fig. 1c is provided when the bias on the detector will exceed ~210 µV.

In order to observe this 'time-resolving' detection, the effect of the detector bias $V_d$ was monitored up to 1.5 mV. Applying a high bias to the detector without modifying the transmission through the QD is by no means a trivial task, since the electrostatic coupling between detector and QD is strong[15]. Hence, the genuine dephasing was excluded from the inadvertent electrostatic effect by setting the detector to a regime where it is not sensitive to the potential in the QD (see Fig. 3b). Compensating for the electrostatic gating effect led to the dependence of the two AB harmonics on detector bias as shown in Fig. 3c. With increasing the detector bias the first-harmonic dropped monotonically, however, the second harmonic remained unaltered in a low bias regime, altering its declining slope at $V_d$~500 µV. Still the suppression rate of the second AB harmonic remained lower than that of the first AB harmonic.

Because of the low transmission probability through the QD, the major contribution of the first harmonic comes from the two direct paths through the interferometer. Thus, the dephasing rate[16-18] of the first AB harmonic can be analyzed in a similar manner to that in the previous experimental work with an open-loop interferometer[5]. The expected visibility (in terms of its fast-Fourier-transformed amplitude $A_1$) has the form, $v=A_1/A_1(V_d=0)=1-\Gamma_d/\Gamma_e$, where $\Gamma_e$ is the natural broadening of the state in the QD due to coupling to the leads, and $\Gamma_d$ is the dephasing rate induced by the charge detection. $\Gamma_e \sim 0.33$ μeV in our experiment[6] and $\Gamma_d$ is given by an algebraic sum of two different contributions $\Gamma_d = \Gamma_T + \Gamma_\emptyset$, where $\Gamma_T$ and $\Gamma_\emptyset$ correspond to the current- and the phase-sensitive dephasing rates, respectively[19-24], expressed as:

$$\Gamma_T = (eV_d/8\pi)(\Delta T_d)^2/T_d(1-T_d),$$

$$\Gamma_\emptyset = (eV_d/2)T_d(1-T_d)(\Delta\emptyset)^2.$$

The phase sensitivity, $\Delta\emptyset$, is defined as the relative phase shift of the transmitted and reflected partial waves induced by an extra charge in the QD. Recently, unexpectedly large dephasing rate was observed and interpreted[15] only in terms of $\Gamma_T$. It has been theoretically proposed that this can be understood by taking into account $\Gamma_\emptyset$ as well, which is much larger than $\Gamma_T$ in a generic situation with a non-negligible asymmetry in the charge response of the QPC potential[21]. In practice, $\Delta\emptyset$ cannot be directly extracted from our measurement setup. Best fitting to the data for the first AB harmonic, with $\Delta\emptyset$ (=0.031) as a fitting parameter, is given by the solid line in Fig. 3c. The fits reveal that the phase-sensitive dephasing mechanism is more effective than that of the current-sensitive dephasing (namely, $\Gamma_\emptyset/\Gamma_T \sim 35$), which is consistent with the previous observation[14,21].

One may still raise the question, why is the dephasing rate of the second AB harmonic at higher detector's bias so low. We speculate that the lower rate may originate from the finite size of electron wave packet in the detector channel, being larger than the interferometer. The time-resolving detection is effective only when the size of the electron wavepackets is infinitesimally small. The dephasing can be alternatively understood in terms of the *back-action*[3], which is the randomization of the phase of an electron passing through the QD due to the fluctuations of the QD potential induced by the current noise in the QPC detector[3].  For the trajectory in Fig. 1b (first-harmonic interference), the random phase is collected for the right path during the entire passage of a wave packet through the QD, while no random phase is collected for the left path, which leads to the suppression of the first-harmonic interference.  However, for the trajectory in Fig. 1c (second-harmonic interference), both wave packets taking the left and the right paths dwell some time in the QD simultaneously and collect *common random phase*, which does not suppress the interference.  The random phase collected while only one of the packets occupies the QD suppresses the interference.  Thus, the random phase accumulated for the second harmonic should be smaller than that of the first harmonic, which gives a qualitative explanation for the lower dephasing rate of the second harmonic.

To date, it has been widely accepted that the quantum mechanical complementarity of a particle can be understood in terms of the momentum transfer (or back action), which is inevitably caused by detecting the path of the particle, as explicitly stated by Feynman[2] a few decades ago.  Recently, however, it has been demonstrated that the particle-like behavior can take place also by the which-path information even for the sufficiently weak momentum transfer[7], which refutes the back-action picture of the dephasing.  In clear contrast to previous works, our work confirms that the wave-like behavior is preserved unless the which-path information can be acquired out of the detection process, even if it can be done only in principle; regardless of the strength of finite 'disturbance' caused by the charge detection.  This has been verified by investigating the second-harmonic dephasing in closed-loop AB-ring interferometer, which has no analogue in the systems studied previously, including the optical[4], solid-state[5], and atomic[7] interferometers.

**Acknowledgement**

This work was supported (for HJL) by Electron Spin Science Center in POSTECH and Pure Basic Research Grant R01-2006-000-11248-0 administered by KOSEF, by the Korea Research Foundation Grant KRF-2005-070-C00055, and by POSTECH Core Research Program. This work was also supported (for YC) by KRISS, KICOS, and NCoE at Hanyang University through a grant provided by the Korean Ministry of Science & Technology, and (for KK) by KRF-2006-331-C00016. MH wishes to acknowledge the partial support of the MINERVA foundation, the German Israeli foundation (GIF), the German Israeli project cooperation (DIP), the Israeli Science foundation (ISF).

**References:**

1. Bohr, N. in *Quantum Theory and Measurement*(eds Wheeler, J. A. & Zurek, W. H.) 9-49 (Princeton Univ. Press, 1983).

2. Feynman, R., Leighton, R. & Sands, M. in The Feynman Lectures on Physics Vol. III, Ch. 1 (Addison Wesley, Reading, 1965).

3. Stern, A., Aharonov, Y. & Imry, Y. Phase uncertainty and loss of interference: A general picture. *Phys. Rev. A* **41**, 3436-3448 (1990).

4. Zou, X. Y., Wang, L. J., & Manel. L. Induced coherence and indistinguishability in optical interference. *Phys. Rev. Lett*, **67**, 318-321 (1991).

5. Buks, E, Schuster, R, Heiblum, M, Mahalu, D, & Umansky, V. Dephasing in electron interference by a 'which-path' detector. *Nature* **391**, 871-874 (1998).

6. Yacoby, A., Heiblum, M., Mahalu, D. & Strikman, H. Coherence and Phase Sensitive Measurements in a Quantum Dot. *Phys. Rev. Lett*. **74**, 4047-4050 (1995).

7. Durr, S., Nonn, T. & Rempe, G. Origin of quantum-mechanical complementarity probed by a 'which-way' experiment in an atom interferometer. *Nature* **395**, 33-37 (1998).

8. Schuster, R. *et al*. Phase measurement in a quantum dot via a double-slit interference experiment. *Nature* **385**, 417-420 (1997).

9. Field, M. *et al*. Measurements of Coulomb blockade with a noninvasive voltage probe. *Phys. Rev. Lett*. **70**, 1311-1314 (1993).

10. Sprinzak, D., Ji, Y., Heiblum, M., Mahalu, D. & Shtrikman, H. Charge Distribution in a Kondo-Correlated Quantum Dot. *Phys. Rev. Lett*. **88**, 176805 (2002).

11. Avinun-Kalish, M., Heiblum, M., Zarchin, O., Mahalu, D. & Umansky, V. Crossover from 'mesoscopic' to 'universal' phase for electron transmission in quantum dots. *Nature* **436**, 529-533 (2005).

12. Khym, G. L. & Kang. K. Charge detection in a closed-loop Aharonov-Bohm interferometer. *Phys. Rev. B*  **74**, 153309 (2006).

13. Aharonov, Y. & Bohm, D. Significance of electromagnetic potentials in the quantum theory. *Phy. Rev*. **115**, 485-491 (1959).

14. Aronov, A. G. & Sharvin, Yu. V. Magnetic flux effects in disordered conductors. *Rev. Mod. Phys*. **59**, 755-779 (1987).

15. Avinun-Kalish, M., Heiblum, M., Silva, A., Mahalu, D. & Umansky, V. Controlled Dephasing of a Quantum Dot in the Kondo Regime. *Phys. Rev. Lett*. **92**, 156801 (2004).

16. Aleiner, I. L., Wingreen, Ned S. & Meir, Y. Dephasing and the Orthogronality Catastrophe in Tunneling through a Quantum Dot: The "Which Path?" Interferometer *Phys. Rev. Lett.* **79**, 3740-3743 (1997).

17. Gurvitz, S. A. Measurements with a noninvasive detector and dephasing mechanism *Phys. Rev. B* **56**, 15215-15223 (1997).

18. Levinson, Y. Dephasing in a quantum dot due to coupling with a quantum point contact. *Europhys. Lett.* **39**, 299-304 (1997).

19. Stodolsky, L. Measurement process in a variable-barrier system. *Phys. Lett. B*. 459, 193-200 (1999).

20. Sprinzak, D., Buks, E., Heiblum, M. & Shtrikman, H. Controlled Dephasing of Electrons *via* a Phase Sensitive Detector. *Phys. Rev. Lett*. **84**, 5820-5823 (2000).

21. Kang. K. Decoherence of the Kondo Singlet via a Quantum Point Contact Detector. *Phys. Rev. Lett*. **95**, 206808 (2005).

22. Buttiker, M. & Martin, A. M. Charge relaxation and dephasing in Coulomb-coupled conductors. *Phys. Rev. B* 61, 2737-2741 (2000).

23. Levinson, Y. Quantum dot dephasing by edge states. *Phys. Rev. B* 61, 4748-4753 (2000).

24. Hackenbroich, G. Phase coherent transmission through interacting mesoscopic systems. *Phys. Rep.* 343, 463-538 (2001).

**Figure Captions:**

**Figure 1. Which-path interferometer**. **a,** SEM image of a closed-loop Aharonov-Bohm (AB) interferometer fabricated on the surface of a two-dimensional electron gas wafer. The device consists of two parts; an electronic interferometer and a quantum point contact (QPC) detector. Four side gates ($M_1$, $M_2$, $M_3$, $M_4$) together with an island gate (P) define an AB ring for the electron transmission. Two pairs of QPC's ($Q_1$, $Q_2$) at the source and the drain were used to configure the two-terminal measurement. A quantum dot (QD), defined by two gates F1 and F2, is embedded in one arm of the interferometer and a QPC is placed in the immediate vicinity of the QD to detect charges trapped in it. The island gate is electrically controlled through the bridged electrode P. **b,** The trajectory leading to the first-harmonic interference ($h/e$ conductance oscillation), where the charge detection is equivalent to the path detection. **c,** The trajectory leading to the second-harmonic interference ($h/2e$ conductance oscillation), where the charge detection is not necessarily the path detection.

**Figure 2. Detection procedure and AB oscillation. a,** With the left path in Fig. 1a pinched off, the Coulomb blockade of the QD and the conductance of the QPC detector are taken by sweeping the island gate voltage ($V_p$). Each time an electron passes through the QD the transmission through the QPC detector is affected, showing a saw-tooth-like behaviour. **b**, With both the left and the right paths open, the coherent transmission of the partial waves leads to AB oscillations.

Around the Coulomb-blockade peak denoted by an arrow in **a** AB oscillations reveal both the first- and the second-harmonic interferences, where the amplitude of the first harmonic is suppressed with increasing the QPC detector bias. But the second-harmonic amplitude is almost insensitive to the detector bias in this low bias range up to 400 μV (see also the inset).

**Figure 3. Time-resolving measurements on the second-harmonic interference. a**, Fast-Fourier-transformed amplitudes of AB oscillations for the first- and the second-harmonic interference with varying the bias of the QPC detector. Inset; schematic diagram illustrating the source-QD transit time difference of electron wave packets between the left ($t_2$) and the right ($t_1$) paths. The source-QD distances of the left and right paths are ~2.5 μm and ~0.83 μm, respectively, which correspond to the transit-time difference of $\Delta t$~10 ps. **b**, The gate voltage ($V_g$) dependence of the QPC transmittance showing the least detection-sensitive region (non-detection regime) and the most-sensitive region (detection regime). **c**, FFT amplitudes for the two harmonics normalized by the corresponding zero-detector-bias values. The amplitude of the first-harmonic interference is suppressed continuously with increasing the detector bias while that of the second-harmonic interference remains almost bias-insensitive in the low bias region below $V_d$~500 μV without the time-resolving power between the left and the right encircling (no which-path information). For $V_d$>500 μV, the detector recovers the time-resolving power (which-path detection), which leads to the suppression of the amplitudes of second harmonic with further increasing $V_d$.

**Figure 1**

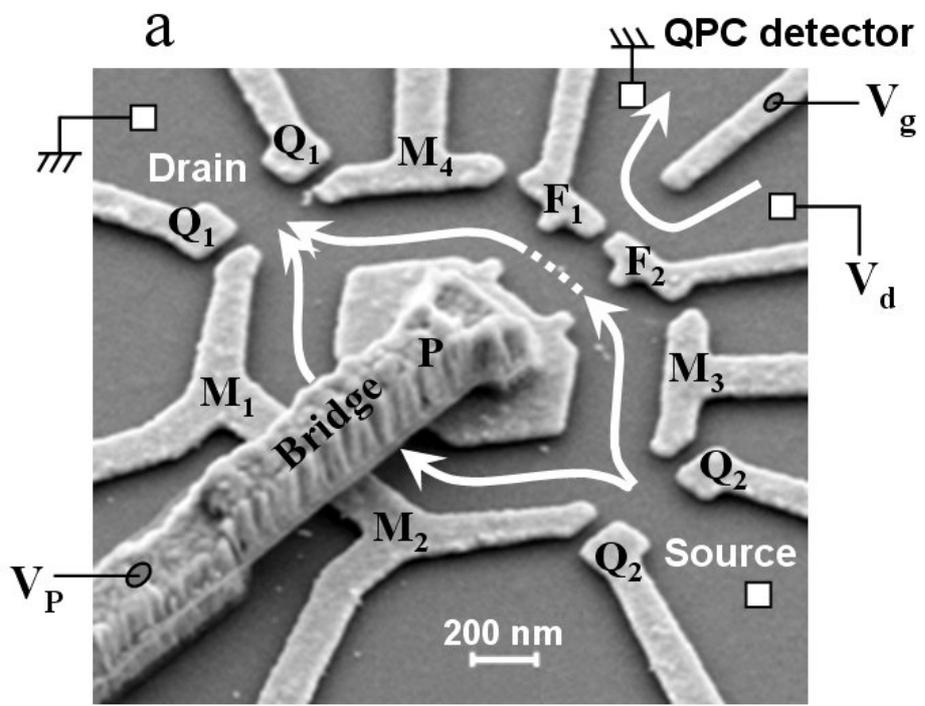

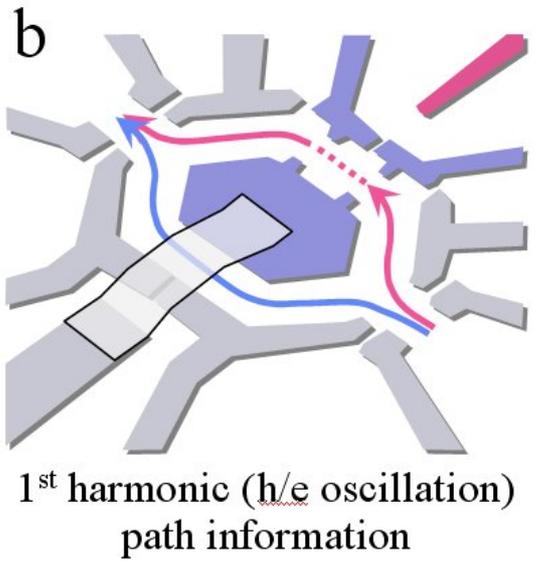

1st harmonic (h/e oscillation)
path information

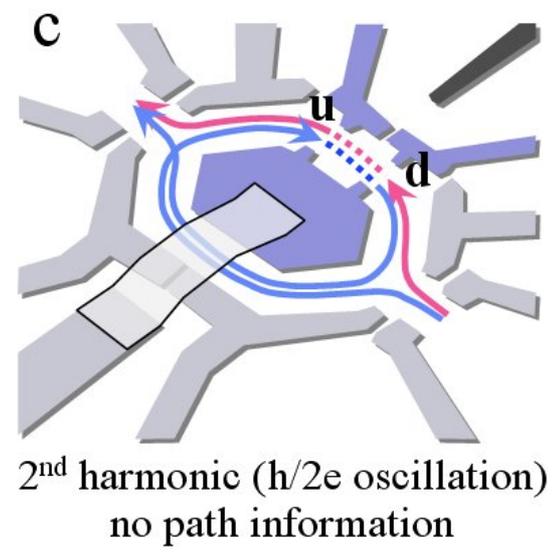

2nd harmonic (h/2e oscillation)
no path information

**Figure 2**

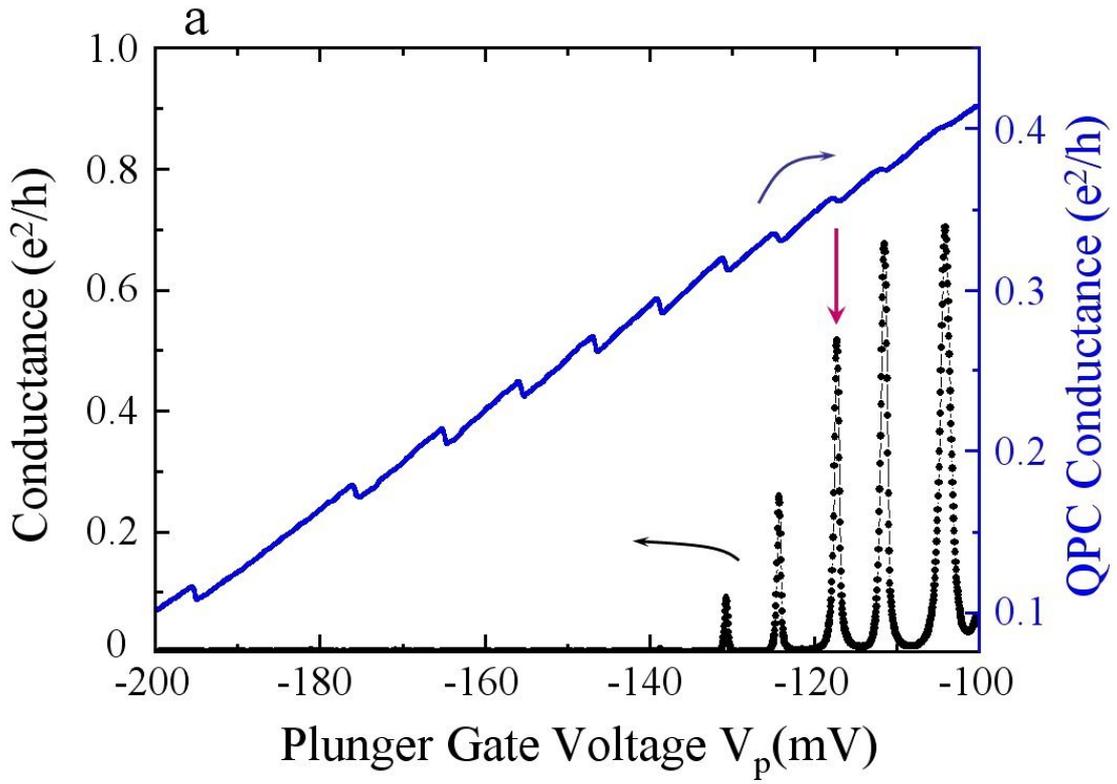

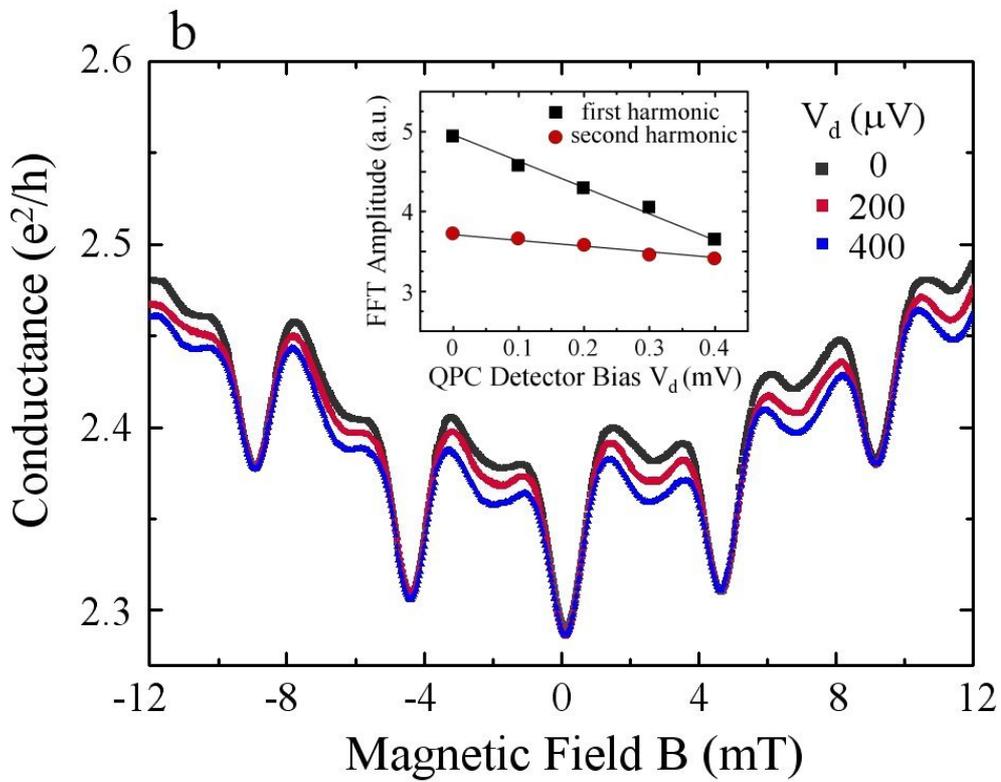

**Figure 3**

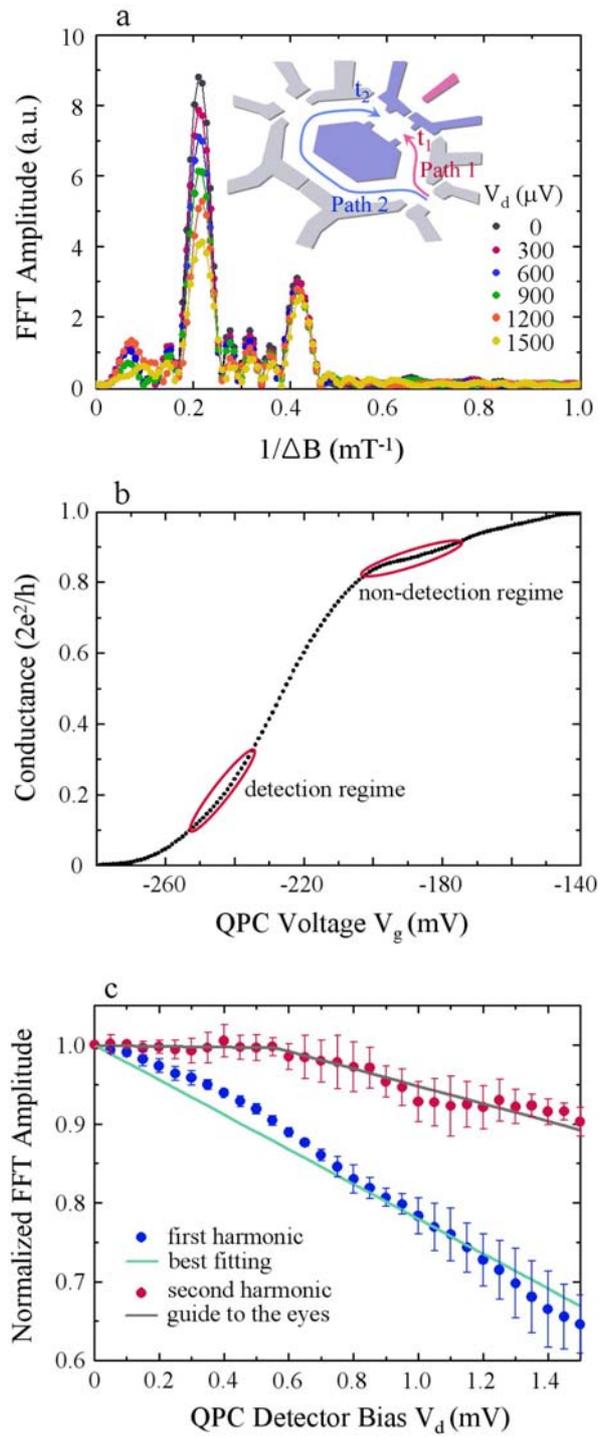